\documentclass[aps,prl,,twocolumn,letterpaper,superscriptaddress]{revtex4}
\usepackage{graphicx}  
\usepackage{tabularx}
\usepackage{bm}        
\usepackage{amssymb}   
\usepackage{subcaption}
\usepackage{amsmath}
\usepackage{threeparttable}
\usepackage{booktabs}
\usepackage{multirow}
\usepackage{siunitx}
\usepackage{mathrsfs,amsfonts}
\usepackage[colorlinks=true]{hyperref} 
\begin{document}
\title{Searching for Two-Neutrino and Neutrinoless Double Beta Decay of $^{134}$Xe with the PandaX-4T Experiment}


\def\shKeyLab{School of Physics and Astronomy, Shanghai Jiao Tong University, Key Laboratory for Particle Astrophysics and Cosmology (MoE), Shanghai Key Laboratory for Particle Physics and Cosmology, Shanghai 200240, China}
\def\BUAA{School of Physics, Beihang University, Beijing 102206, China}
\def\BUAALab{Beijing Key Laboratory of Advanced Nuclear Materials and Physics, Beihang University, Beijing, 102206, China}
\def\zzu{School of Physics and Microelectronics, Zhengzhou University, Zhengzhou, Henan 450001, China}
\def\USTClab{State Key Laboratory of Particle Detection and Electronics, University of Science and Technology of China, Hefei 230026, China}
\def\USTCdep{Department of Modern Physics, University of Science and Technology of China, Hefei 230026, China}
\def\BUAALab{International Research Center for Nuclei and Particles in the Cosmos \& Beijing Key Laboratory of Advanced Nuclear Materials and Physics, Beihang University, Beijing 100191, China}
\def\pku{School of Physics, Peking University, Beijing 100871, China}
\def\YaLongSD{Yalong River Hydropower Development Company, Ltd., 288 Shuanglin Road, Chengdu 610051, China}
\def\IAP{Shanghai Institute of Applied Physics, Chinese Academy of Sciences, 201800 Shanghai, China}
\def\CHEPpku{Center for High Energy Physics, Peking University, Beijing 100871, China}
\def\SDUdep{Research Center for Particle Science and Technology, Institute of Frontier and Interdisciplinary Science, Shandong University, Qingdao 266237, Shandong, China}
\def\SDUlab{Key Laboratory of Particle Physics and Particle Irradiation of Ministry of Education, Shandong University, Qingdao 266237, Shandong, China}
\def\UMD{Department of Physics, University of Maryland, College Park, Maryland 20742, USA}
\def\TDLee{Tsung-Dao Lee Institute, Shanghai Jiao Tong University, Shanghai, 200240, China}
\def\MESJTU{School of Mechanical Engineering, Shanghai Jiao Tong University, Shanghai 200240, China}
\def\SYU{School of Physics, Sun Yat-Sen University, Guangzhou 510275, China}
\def\SYUSFI{Sino-French Institute of Nuclear Engineering and Technology, Sun Yat-Sen University, Zhuhai, 519082, China}
\def\NKU{School of Physics, Nankai University, Tianjin 300071, China}
\def\YTU{Department of Physics,Yantai University, Yantai 264005, China}
\def\FDU{Key Laboratory of Nuclear Physics and Ion-beam Application (MOE), Institute of Modern Physics, Fudan University, Shanghai 200433, China}
\def\USST{School of Medical Instrument and Food Engineering, University of Shanghai for Science and Technology, Shanghai 200093, China}
\def\SJTUSC{Shanghai Jiao Tong University Sichuan Research Institute, Chengdu 610213, China}
\def\SPEIT{SJTU Paris Elite Institute of Technology, Shanghai Jiao Tong University, Shanghai, 200240, China}
\def\NNU{School of Physics and Technology, Nanjing Normal University, Nanjing 210023, China}
\def\SYUzhuhai{School of Physics and Astronomy, Sun Yat-Sen University, Zhuhai, 519082, China}

\affiliation{\shKeyLab}
\author{Xiyu Yan}\affiliation{\SYUzhuhai}
\author{Zhaokan Cheng}\affiliation{\SYUSFI}
\author{Abdusalam Abdukerim}\affiliation{\shKeyLab}
\author{Zihao Bo}\affiliation{\shKeyLab}
\author{Wei Chen}\affiliation{\shKeyLab}
\author{Xun Chen}\affiliation{\shKeyLab}\affiliation{\SJTUSC}
\author{Chen Cheng}\affiliation{\SYU}

\author{Xiangyi Cui}\affiliation{\TDLee}
\author{Yingjie Fan}\affiliation{\YTU}
\author{Deqing Fang}\affiliation{\FDU}
\author{Changbo Fu}\affiliation{\FDU}
\author{Mengting Fu}\affiliation{\pku}
\author{Lisheng Geng}\affiliation{\BUAA}\affiliation{\BUAALab}\affiliation{\zzu}
\author{Karl Giboni}\affiliation{\shKeyLab}
\author{Linhui Gu}\affiliation{\shKeyLab}
\author{Xuyuan Guo}\affiliation{\YaLongSD}
\author{Chencheng Han}\affiliation{\TDLee} 
\author{Ke Han}\email[Corresponding author: ]{ke.han@sjtu.edu.cn}\affiliation{\shKeyLab}
\author{Changda He}\affiliation{\shKeyLab}
\author{Jinrong He}\affiliation{\YaLongSD}
\author{Di Huang}\affiliation{\shKeyLab}
\author{Yanlin Huang}\affiliation{\USST}
\author{Junting Huang}\affiliation{\shKeyLab}
\author{Zhou Huang}\affiliation{\shKeyLab}
\author{Ruquan Hou}\affiliation{\SJTUSC}
\author{Yu Hou}\affiliation{\MESJTU}
\author{Xiangdong Ji}\affiliation{\UMD}
\author{Yonglin Ju}\affiliation{\MESJTU}
\author{Chenxiang Li}\affiliation{\shKeyLab}
\author{Jiafu Li}\affiliation{\SYU}
\author{Mingchuan Li}\affiliation{\YaLongSD}
\author{Shuaijie Li}\affiliation{\TDLee}
\author{Tao Li}\affiliation{\SYUSFI}
\author{Qing Lin}\affiliation{\USTClab}\affiliation{\USTCdep}
\author{Jianglai Liu}\email[Spokesperson: ]{jianglai.liu@sjtu.edu.cn}\affiliation{\shKeyLab}\affiliation{\TDLee}\affiliation{\SJTUSC}
\author{Xiaoying Lu}\affiliation{\SDUdep}\affiliation{\SDUlab}
\author{Congcong Lu}\affiliation{\MESJTU}
\author{Lingyin Luo}\affiliation{\pku}
\author{Yunyang Luo}\affiliation{\USTCdep}
\author{Wenbo Ma}\affiliation{\shKeyLab}
\author{Yugang Ma}\affiliation{\FDU}
\author{Yajun Mao}\affiliation{\pku}
\author{Yue Meng}\affiliation{\shKeyLab}\affiliation{\SJTUSC}
\author{Xuyang Ning}\affiliation{\shKeyLab}
\author{Binyu Pang}\affiliation{\SDUdep}\affiliation{\SDUlab}
\author{Ningchun Qi}\affiliation{\YaLongSD}
\author{Zhicheng Qian}\affiliation{\shKeyLab}
\author{Xiangxiang Ren}\affiliation{\SDUdep}\affiliation{\SDUlab}
\author{Nasir Shaheed}\affiliation{\SDUdep}\affiliation{\SDUlab}
\author{Xiaofeng Shang}\affiliation{\shKeyLab}
\author{Xiyuan Shao}\affiliation{\NKU}
\author{Guofang Shen}\affiliation{\BUAA}
\author{Lin Si}\affiliation{\shKeyLab}
\author{Wenliang Sun}\affiliation{\YaLongSD}
\author{Andi Tan}\affiliation{\UMD}
\author{Yi Tao}\affiliation{\shKeyLab}\affiliation{\SJTUSC}
\author{Anqing Wang}\affiliation{\SDUdep}\affiliation{\SDUlab}
\author{Meng Wang}\affiliation{\SDUdep}\affiliation{\SDUlab}
\author{Qiuhong Wang}\affiliation{\FDU}
\author{Shaobo Wang}\email[Corresponding author: ]{shaobo.wang@sjtu.edu.cn}\affiliation{\shKeyLab}\affiliation{\SPEIT}
\author{Siguang Wang}\affiliation{\pku}
\author{Wei Wang}\affiliation{\SYUSFI}\affiliation{\SYU}
\author{Xiuli Wang}\affiliation{\MESJTU}
\author{Zhou Wang}\affiliation{\shKeyLab}\affiliation{\SJTUSC}\affiliation{\TDLee}
\author{Yuehuan Wei}\affiliation{\SYUSFI}
\author{Mengmeng Wu}\affiliation{\SYU}
\author{Weihao Wu}\affiliation{\shKeyLab}
\author{Jingkai Xia}\affiliation{\shKeyLab}
\author{Mengjiao Xiao}\affiliation{\UMD}
\author{Xiang Xiao}\affiliation{\SYU}
\author{Pengwei Xie}\affiliation{\TDLee}
\author{Binbin Yan}\affiliation{\TDLee}
\author{Jijun Yang}\affiliation{\shKeyLab}
\author{Yong Yang}\affiliation{\shKeyLab}
\author{Yukun Yao}\affiliation{\shKeyLab}
\author{Chunxu Yu}\affiliation{\NKU}
\author{Ying Yuan}\affiliation{\shKeyLab}
\author{Zhe Yuan}\affiliation{\FDU} %
\author{Xinning Zeng}\affiliation{\shKeyLab}
\author{Dan Zhang}\affiliation{\UMD}
\author{Minzhen Zhang}\affiliation{\shKeyLab}
\author{Peng Zhang}\affiliation{\YaLongSD}
\author{Shibo Zhang}\affiliation{\shKeyLab}
\author{Shu Zhang}\affiliation{\SYU}
\author{Tao Zhang}\affiliation{\shKeyLab}
\author{Wei Zhang}\affiliation{\TDLee}
\author{Yang Zhang}\affiliation{\SDUdep}\affiliation{\SDUlab}
\author{Yingxin Zhang}\affiliation{\SDUdep}\affiliation{\SDUlab} %
\author{Yuanyuan Zhang}\affiliation{\TDLee}
\author{Li Zhao}\affiliation{\shKeyLab}
\author{Qibin Zheng}\affiliation{\USST}
\author{Jifang Zhou}\affiliation{\YaLongSD}
\author{Ning Zhou}\affiliation{\shKeyLab}\affiliation{\SJTUSC}
\author{Xiaopeng Zhou}\affiliation{\BUAA}
\author{Yong Zhou}\affiliation{\YaLongSD}
\author{Yubo Zhou}\affiliation{\shKeyLab}
\collaboration{PandaX Collaboration}
\noaffiliation
\date{\today}
\begin{abstract}
$^{134}$Xe is a candidate isotope for neutrinoless double beta decay~($0\nu\beta\beta$) search. 
In addition, the two-neutrino case ($2\nu\beta\beta$) allowed by the standard model of particle physics has not yet been observed.
With the 656-kg natural xenon in the fiducial volume of the PandaX-4T detector, which contains 10.4\% of $^{134}$Xe, and its initial 94.9-day exposure, we have established the most stringent constraints on $2\nu\beta\beta$ and $0\nu\beta\beta$ of $^{134}$Xe half-lives, with limits of $2.8\times10^{22}$~yr and $3.0\times10^{23}$~yr at 90\% confidence level, respectively. 
The $2\nu\beta\beta$ ($0\nu\beta\beta$) limit surpasses the previously reported best result by a factor of 32 (2.7), highlighting the potential of large monolithic natural xenon detectors for double beta decay searches.

\end{abstract}
\maketitle

Neutrinoless double beta decay ($0\nu\beta\beta$), if discovered, would be the unambiguous evidence of the Majorana nature of neutrino and the violation of lepton number conservation~\cite{Furry:1939qr, Elliott:2014iha, Dolinski:2019nrj}.
$0\nu\beta\beta$ could only happen in isotopes decaying via two-neutrino double beta decay~($2\nu\beta\beta$), an exceptionally rare nuclear decay process emitting two electrons and two antineutrinos simultaneously~\cite{Goeppert-Mayer:1935uil, Avignone:2007fu}. 
$2\nu\beta\beta$ has been directly observed in multiple isotopes, with the longest half-life at approximately $2\times 10^{21}$~yr for $^{136}$Xe~\cite{EXO-200:2013xfn, KamLAND-Zen:2019imh, NEXT:2021dqj, PandaX:2022kwg}.
The $2\nu\beta\beta$ of $^{134}$Xe, with an expected half-life of $\mathcal{O}(10^{24})$~yr~\cite{Staudt:1990qi,Barros:2014exa}, is the next promising discovery. 
The most stringent experimental constraint on its half-life is $8.7 \times 10^{20}$~yr at the 90\% confidence level (CL) from the EXO-200 experiment~\cite{EXO-200:2017vqi}, which used xenon with approximately 19\% $^{134}$Xe (a fiducial mass of 18.1~kg) and 81\% $^{136}$Xe.

Different from the $2\nu\beta\beta$, $0\nu\beta\beta$ has no neutrinos emitted.
Many experiments have searched for the $0\nu\beta\beta$ processes employing various isotopes, including $^{76}$Ge~\cite{GERDA:2020xhi,Majorana:2022udl}, $^{130}$Te~\cite{CUORE:2022jto}, and $^{136}$Xe~\cite{EXO-200:2019rkq, KamLAND-Zen:2022tow}. The best constraints on the half-lives of these isotopes are on the order of $10^{26}$~yr.
EXO-200 has reported the most stringent constraint on the $0\nu\beta\beta$ half-life of $^{134}$Xe, $1.1 \times 10^{23}$~yr at the 90\% CL~\cite{EXO-200:2017vqi}.
If discoveries are made in a pair of isotopes of the same element, such as $^{136}$Xe and $^{134}$Xe, it is particularly intriguing to potentially mitigate theoretical uncertainties associated with alternative $0\nu\beta\beta$ mechanisms, such as heavy Majorana neutrino exchange or gluino exchange in supersymmetry models~\cite{LZ:2021blo}.

Previously, we have reported a precise measurement of the $2\nu\beta\beta$ half-life of $^{136}$Xe~\cite{PandaX:2022kwg}.
In this Letter, we present new searches for $^{134}$Xe $2\nu\beta\beta$ and $0\nu\beta\beta$ using the PandaX-4T natural xenon Time Projection Chamber (TPC)~\cite{PandaX-4T:2021bab} based on 94.9 live days of data from the initial data release.
Compared to searches conducted in a $^{136}$Xe-enriched detector, a natural xenon detector such as PandaX-4T offers the unique advantage of a high $^{134}$Xe/$^{136}$Xe ratio of 1.2.
The energy region of interest~(ROI) from 200 to 1000~keV includes the majority of the $2\nu\beta\beta$ events and almost all of the $0\nu\beta\beta$ events.
Compared to EXO-200, our extension of the energy range to 200~keV also results in a relatively lower contribution from $^{136}$Xe $2\nu\beta\beta$ due to the different Q values. 

The PandaX-4T experiment is in the B2 hall of the recently expanded China Jinping Underground Laboratory~\cite{Li:2014rca}.
The detector has been commissioned since late 2020.
The cylindrical active volume of the PandaX-4T dual-phase TPC, operating at $-94.6~^\circ$C, is enclosed within a field cage to ensure a uniform electric field along the vertical $z$ axis.
The active volume has a diameter of 118.5~cm and a height of 116.8~cm.
The height is calculated as the distance between the cathode and gate meshes, considering the 1.4\% polytetrafluoroethylene shrinkage caused by temperature.
As a result, the total mass of natural xenon in the active volume is 3.69 tonnes, slightly different from the previously reported value~\cite{PandaX:2022kwg}.
The TPC is equipped with 169 and 199 three-inch Hamamatsu photomultiplier tubes~(PMTs) mounted at the top and the bottom readout arrays in horizontal $xy$ direction, respectively. For more details about the detector configuration, please refer to Refs.~\cite{PandaX-4T:2021bab, PandaX:2022kwg}.

PandaX-4T has developed a dedicated analysis pipeline for MeV-scale signals and performed calibration campaigns in the energy range, allowing for the study of double beta decays of $^{136}$Xe~\cite{PandaX:2022kwg}.
Following a similar procedure, the search for the $2\nu\beta\beta$ and $0\nu\beta\beta$ of $^{134}$Xe was performed with the same data set and fiducial volume (FV).
Energy deposition in liquid xenon (LXe) produces prompt scintillation light $S1$ signals and ionization electrons.
The ionization electrons drift toward the liquid-gas interface in the applied electric drift field.
The electric field in the gaseous region is more than 1 order of magnitude stronger at 5--6 kV/cm, resulting in the emission of $S2$ photons through electroluminescence. 
The distribution of $S2$ photons, collected by the top PMT array~($S2{_T}$), is utilized to reconstruct the horizontal $xy$ position.
The time delay between the $S1$ and $S2$ signals determines the vertical $z$ position.
Event energy is determined by combining the $S1$ and $S2$ detected by the bottom PMT array~($S2{_B}$).

One improvement in this analysis is the usage of the ``rolling gain" of PMTs, which is determined based on the single photoelectron peaks in physics data. 
PMT gains were calculated with weekly calibration using light-emitting diodes (LEDs) in previous analyses.
The rolling-gain method does not suffer from uneven illumination and has large samples of single photoelectron, which improves the accuracy of gain calibration. 
The gains of all PMTs calculated from the LEDs exhibit an average offset of +1.7\% with a 4.5\% variation compared to the rolling-gain results. 
The difference introduces a negligible difference in the final spectrum but may impact the energy and and position reconstruction of individual events. 
The event selection and spectrum fit procedures are developed with LED-gain data and then applied to the analysis of rolling-gain data.
The stepwise approach minimizes possible bias in the analysis.

In this analysis, we also improve the photon-acceptance-function-based position reconstruction algorithm, which characterizes the anticipated charge detected by a specific PMT as a function of the event's position~\cite{PANDA-X:2021jua}.
The $xy$ position of an event is determined by comparing the observed $S2{_T}$ charge distribution with photon acceptance functions derived from optical Monte Carlo~(MC) simulation.
Compared to the previous analysis~\cite{PandaX:2022kwg}, we use a more realistic model with multiple layers of image PMTs to simulate light reflection on side polytetrafluoroethylene panels.
The $S2{_T}$ charge is calculated with the desaturation algorithm to mitigate PMT saturation of large signals.
Both improvements lead to a more accurate position reconstruction in the $xy$ plane. 
The boundary of the detector is reconstructed within 1.8~cm of the physical position, while an inward offset of 6.0~cm is observed in the previous analysis.

\begin{figure}[tb]
    \includegraphics[width=\columnwidth]{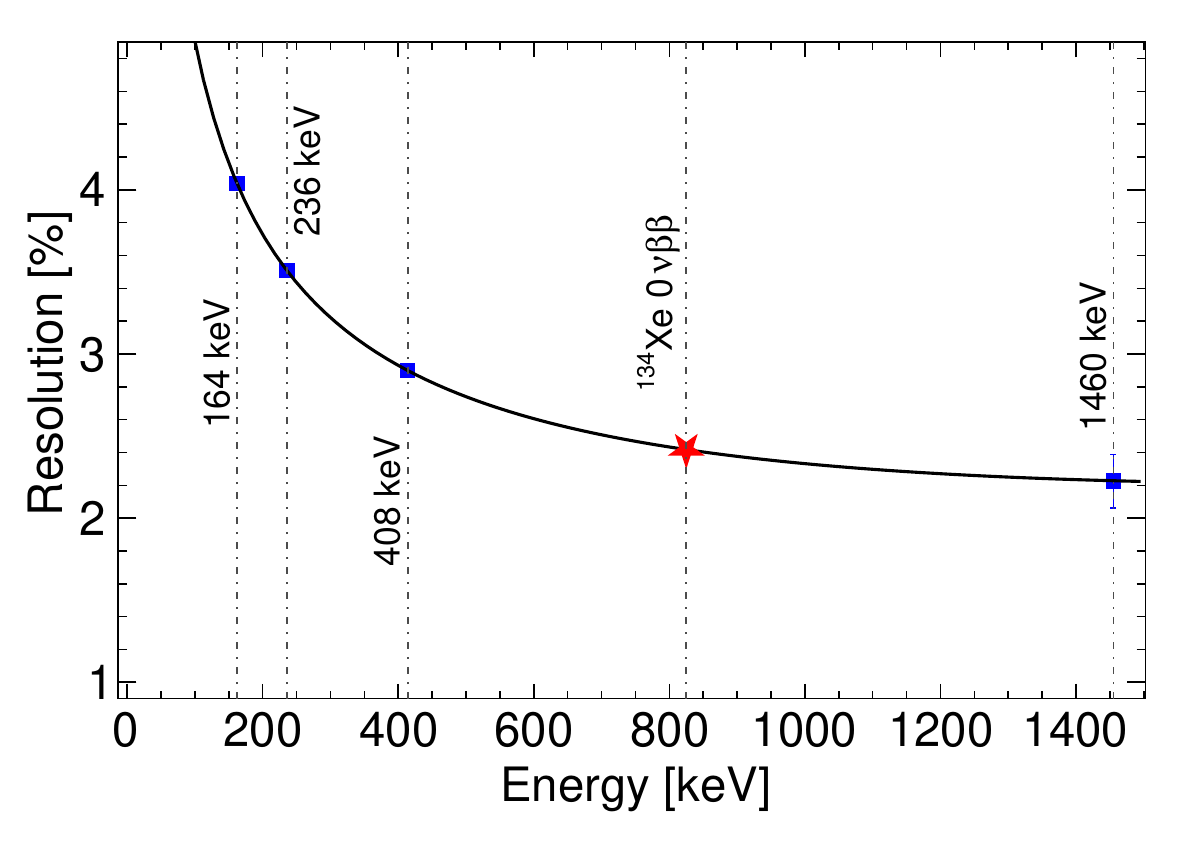}
    \caption{Relative energy resolution $\sigma$/E as a function of energy. 
    The black line represents a fit of the experimental data with the function $0.41/\sqrt{\mathrm{E~[keV]}} + 2.6\times 10^{-6}\mathrm{E~[keV]} + 0.008$.}
    \label{fig:energy reconstruction}
\end{figure}

The detector is calibrated with gamma sources deployed outside the active volume (including $^{232}$Th, $^{60}$Co, and $^{137}$Cs) and internal sources (including $^{\mathrm{83m}}$Kr, $^{\mathrm{131m}}$Xe, $^{127}$Xe, and $^{\mathrm{129m}}$Xe). 
Energy is calculated from the $S1$ signal and $S2{_B}$ signal:
\begin{equation}
 E  = 13.7~\mathrm{eV}  \times \left(\frac{S1}{\mathrm{PDE}}+\frac{S2_B}{\mathrm{EEE \times SEG_B}}\right),
\end{equation}
in which $\mathrm{PDE}$, $\mathrm{EEE}$, and $\mathrm{SEG{_B}}$ are the photon detection efficiency for $S1$, electron extraction efficiency, and the single-electron gain for $S2{_B}$, respectively.
The 13.7~eV is the work function in liquid xenon~\cite{Szydagis:2011tk}.
The $S2_B$ signal is desaturated following the previous analysis~\cite{PandaX:2022kwg}.

The nonuniformity of detector response is corrected in three dimensions.
Along the $z$ direction, electron lifetime correction is applied to $S2{_B}$ to correct electron loss along the drift path. 
The electron lifetime measured with the 164 keV energy peak from $^{\mathrm{131m}}$Xe spans from 800.4 to 1288.2~$\mu$s, depending on the xenon purity~\cite{PandaX-4T:2021bab}. 
The nonuniformity of detector response is mapped out by injecting $^{\mathrm{83m}}$Kr~\cite{Zhang:2021shp} with the improved position reconstruction.
Nonuniformity in the $xyz$ ($xy$) space for $S1$ ($S2{_B}$) is mapped out, and corresponding corrections are applied.
Additional third-order polynomial-function corrections are applied to the energy spectrum based on observed gamma peaks to account for the residual nonlinearity after the position-dependent correction. 
The polynomial correction reduces deviations of gamma peak energies from about 40~keV to less than 10~keV, which is taken as the systematic uncertainty of energy reconstruction.

Figure.~\ref{fig:energy reconstruction} shows the energy resolution of physics data in the FV and is fitted with the parametrization function $a/\sqrt{\mathrm{E}} + b\mathrm{E} + c$. 
The energy resolution reaches 2.4\% for single-site~(SS) events at 825.8~keV, the $Q$ value of $^{134}$Xe double beta decay.
The signal response model is used to generate probability density functions~(PDFs) of signals and backgrounds in the MC.

\begin{figure}[tb]
    \includegraphics[width=\columnwidth]{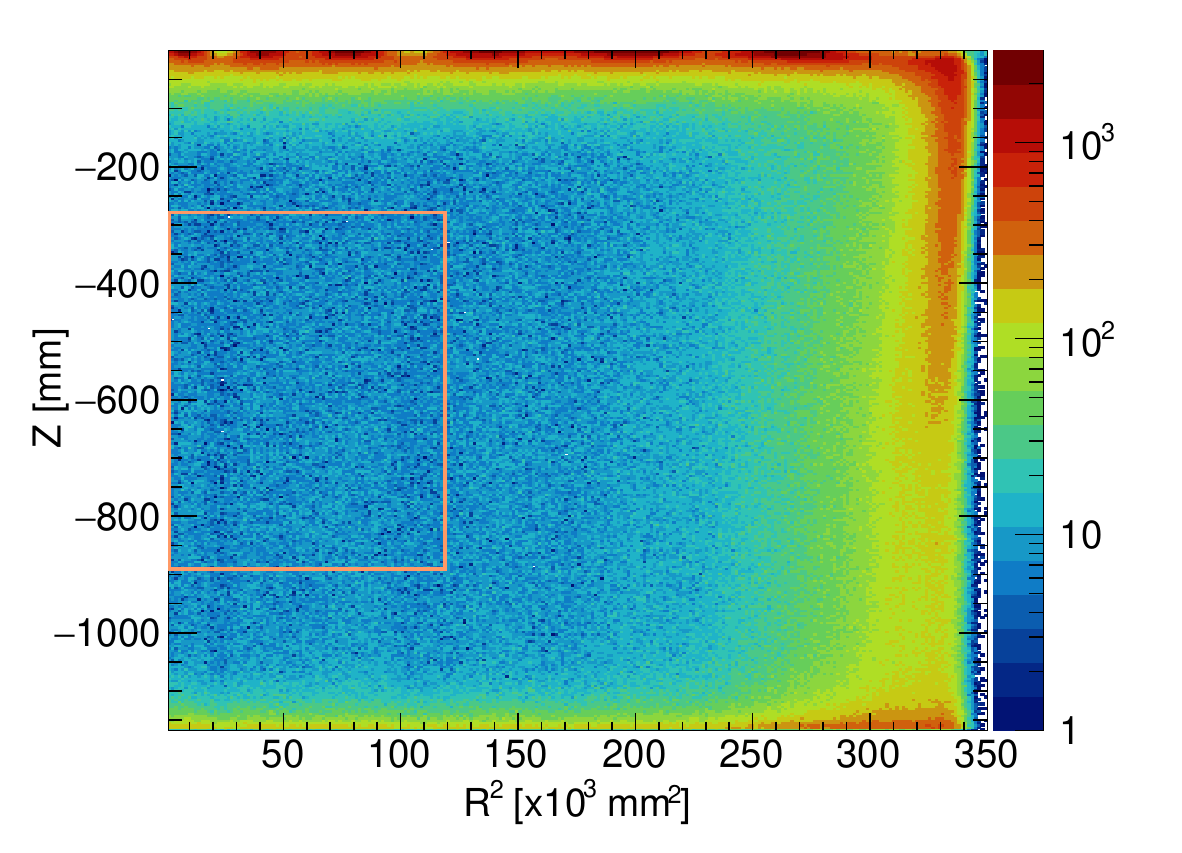}
    \caption{Event distributions of the physics data in the energy range of [200, 1000]~keV in $z$ vs. R$^2$ coordinates, 
    with the color bar indicating counts in each bin. The orange rectangle represents the cylindrical FV.}
    \label{fig:FV}
\end{figure}

The FV, as shown in Fig.~\ref{fig:FV}, is defined by a cylinder with a radius of 34.6~cm and a height of 61.2~cm at the geometrical center of the detector. 
The mass of natural xenon within the FV is 656~kg, calculated by scaling $^{\mathrm{83m}}$Kr percentage and improved position reconstruction.
Taking into account the isotopic abundance of 10.47\%, the FV contains 68.7~kg of $^{134}$Xe.
The total isotopic exposure is 17.9~kg yr.

The criteria for selecting physical events are developed with calibration data, validated with 9.6 days of rolling-gain data, and then applied to the entire physics dataset.
Quality cut variables to remove noise and select electron recoil events are adopted from the previous analysis~\cite{PandaX:2022kwg} but adjusted according to events extended down to 200~keV.
The overall cut efficiency is $(99.94\pm0.03)\%$, and consistent results are calculated from calibration data and the 9.6-day dataset.

The discrimination between SS and multisite (MS) events is improved in this analysis.
Most of the $2\nu\beta\beta$ and $0\nu\beta\beta$ signals are SS events, which can be discriminated from MS gamma background events using the temporal profile of the summed $S2{_B}$ signal.
Energy depositions at different positions along the $z$ axis correspond to different timings in the $S2{_B}$ signal.   
Because of electron diffusion during drifting, the $S2{_B}$ signal is broadened, and the $z$ position information is smeared. 
Instead of a constant time window, $S2{_B}$ peaks separated by more than a dynamic time window ranging from $\pm$0.16 to $\pm$1.6~$\mu$s, are classified as MS events. 
The width of the dynamic window depends on the drift distance and is determined by the $1\sigma$ width of the highest $S2{_B}$ waveform.

\begin{figure}[tb]
    \includegraphics[width=\columnwidth]{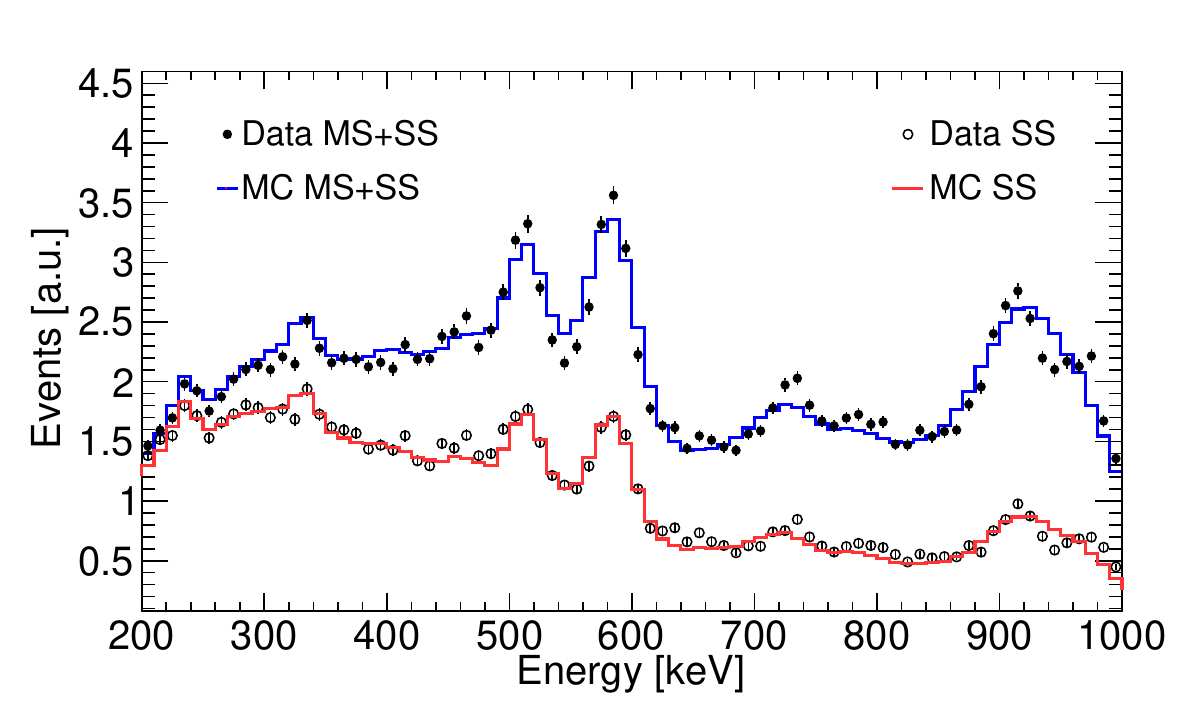}
    \caption{Energy spectra of $^{232}$Th  calibration data and MC simulation for MS~+~SS and SS events. 
    }
    \label{fig:SSMS}
\end{figure}

The SS/(SS+MS) ratio for $2\nu\beta\beta$ ($0\nu\beta\beta$) events is 99.89\% (98.23\%) within the ROI.
The efficiencies are calculated based on BambooMC, a Geant4-based MC framework developed by PandaX~\cite{Chen:2021asx}. 
The $^{134}$Xe $2\nu\beta\beta$ and $0\nu\beta\beta$ events are simulated based on theoretical calculations described in Ref.~\cite{Kotila:2013gea}. 
The SS signal and background PDFs are generated from MC data with the SS criteria identical to those in the physics data. 
The $^{232}$Th gamma calibration data are used to study the systematic uncertainty of SS cut efficiency, as illustrated in Fig.~\ref{fig:SSMS}. 
The difference in SS/(SS+MS) ratio between data and MC (in absolute value) averaged over the energy range is 6.4\%, which is conservatively adopted as a systematic uncertainty for all PDFs.
Since the majority of events in the ROI are $\beta$-like events, which are almost entirely SS events, the systematic uncertainty from $^{232}$Th calibration is a conservative choice.

We search for $^{134}$Xe $2\nu\beta\beta$ and $0\nu\beta\beta$ by fitting the energy spectrum of the SS events with a binned likelihood method.
Similar to Ref.~\cite{PandaX:2022kwg}, the likelihood function is constructed as
\begin{equation}
\begin{aligned}
    L = \displaystyle\prod_{i = 1}^{N_{\rm{bins}}}\frac{(N_{i})^{N_{i,\rm{obs}}}}{N_{i,\rm{obs}}!}e^{-N_{i}} 
    \cdot \prod_{j = 1}^{N_{\rm{G}}} G(0,\sigma_j) \cdot \mathscr{G}(a,b,c, \sigma_{a,b,c})\,,
\label{eq:PL1}
\end{aligned}
\end{equation}
where $N_{i,\rm{obs}}$ and $N_{i}$ are the observed and expected numbers of events in the $i$ th energy bin, with
\begin{equation}
    N_{i} = (1+\epsilon_{\rm{eff}}) \left[(1+\eta_{\rm{Xe}}) \sum_{s}^{2\nu,0\nu} n_s S_{s,i}+\sum_{b=1}^{N_{\rm{bkgs}}}(1+\eta_{b})n_b B_{b,i}\right]\,.
\label{eq:PL2}
\end{equation}
Here, $n_s$ and $n_{b}$ are the counts for the signal ($2\nu\beta\beta$ or $0\nu\beta\beta$) and the background component $b$, respectively, and $S_{s,i}$ and $B_{b,i}$ are the corresponding PDF values in the $i$ th bin. $\epsilon_{\rm{eff}}$, $\eta_{\rm{Xe}}$, and $\eta_b$ are the constrained parameters to represent the systematic uncertainties from the overall efficiency, xenon target mass, and different background contributions.
The constraints are implemented as Gaussian penalty terms in the likelihood function in Eq.(\ref{eq:PL1}) with the 1$\sigma$ variation defined in the text.
The shapes of signal and background PDFs are modeled with energy resolution function with the parameters $a$, $b$, and $c$ (Fig.~\ref{fig:energy reconstruction}) with their correlated uncertainties included in the penalty term $\mathscr{G}(a,b,c, \sigma_{a,b,c})$.

The total detection efficiencies for $2\nu\beta\beta$ ($0\nu\beta\beta$) signals are the product of the data quality efficiency 99.94\% (99.94\%), SS cut efficiency 99.89\% (98.23\%), and ROI acceptance 60.56\% (99.98\%).

The contribution of $^{136}$Xe $2\nu\beta\beta$ background is calculated based on the half-life measurement in Ref.~\cite{PandaX:2022kwg}. 
The number of $^{136}$Xe $2\nu\beta\beta$ events in the ROI is expected to be 12 $372\pm619$, used as the initial value and constraint.
The spectral shape of $^{136}$Xe $2\nu\beta\beta$  events is generated with the Decay0 package~\cite{Ponkratenko:2000um}. $\mathrm{^{133}Xe}$ is generated through the deexcitation of $\mathrm{^{133m}Xe}$, which is the product of the neutron capture process of $\mathrm{^{132}Xe}$.
The number of $^{133}$Xe events is estimated to be $3423\pm342$, based on the data between 90 and 120keV, where the $\beta+\gamma$ shoulder of $^{133}$Xe dominates.
Additional contributions from short-lived xenon isotopes induced by neutron calibration are mainly from characteristic peaks in the data, including 248~keV and 276~keV from $\mathrm{^{125}Xe}$, 208~keV, 380~keV, and 408~keV from $\mathrm{^{127}Xe}$, and 236~keV from $\mathrm{^{129m}Xe}$ and $\mathrm{^{127}Xe}$.

$^{85}$Kr concentration is measured based on a sequential $\beta-\gamma$ emission through the metastable state $^{\mathrm{85m}}$Rb. 
In comparison to Ref.~\cite{PandaX-4T:2021bab}, we optimize the coincidence selection, and the updated number of $^{\mathrm{85m}}$Rb events is $3.9\pm 2.0$, after the subtraction of possible random coincidence.
The concentration of Kr/Xe is $0.52 \pm 0.27$ parts per trillion, assuming an isotopic abundance of $2\times 10^{-11}$ for $^{85}$Kr~\cite{Collon:2004xs}.
A total of $296\pm 154$ $^{85}$Kr events is expected in the ROI.

$\beta$ decays of $^{212}$Pb ($^{214}$Pb), progeny of emanated $^{220}$Rn ($^{222}$Rn), are inferred from data with energy spectra taken from MC. 
Because of the short 55.6-s half-life of $^{220}$Rn~\cite{Wu:2007cdl}, the alpha rate of $^{220}$Rn in the detector does not reflect the internal activity of $^{212}$Pb. 
The $^{212}$Pb activity is instead constrained with daughter $^{212}$Po alpha decay, which is $0.11\pm 0.01~\mu$Bq/kg.
The ratio between $^{212}$Pb/$^{212}$Po of $2.5\pm0.8$ is determined from $^{220}$Rn calibration data. 
Therefore, the nominal activity of $^{212}$Pb is $0.28\pm 0.08 ~\mu$Bq/kg, resulting in $1012 \pm 289$ events in the ROI. 
The $^{222}$Rn level is measured with the alpha particles emitted in the decay chain and is higher than that of $^{214}$Pb due to progenies attaching to electrodes and the inner surface of the detector~\cite{Ma:2020kll}. 
Therefore, the rate of $^{214}$Pb is float in the fit. 

External background originates from the detector components and the lab environment.
The corresponding activities of $^{232}$Th, $^{238}$U, $^{60}$Co, and $^{40}$K have been measured in Ref.~\cite{PandaX:2022kwg} and used as nominal values for the fit.

\begin{figure}[tb]
    \renewcommand{\figurename}{Fig.}
    \centering
 \includegraphics[width=\linewidth]{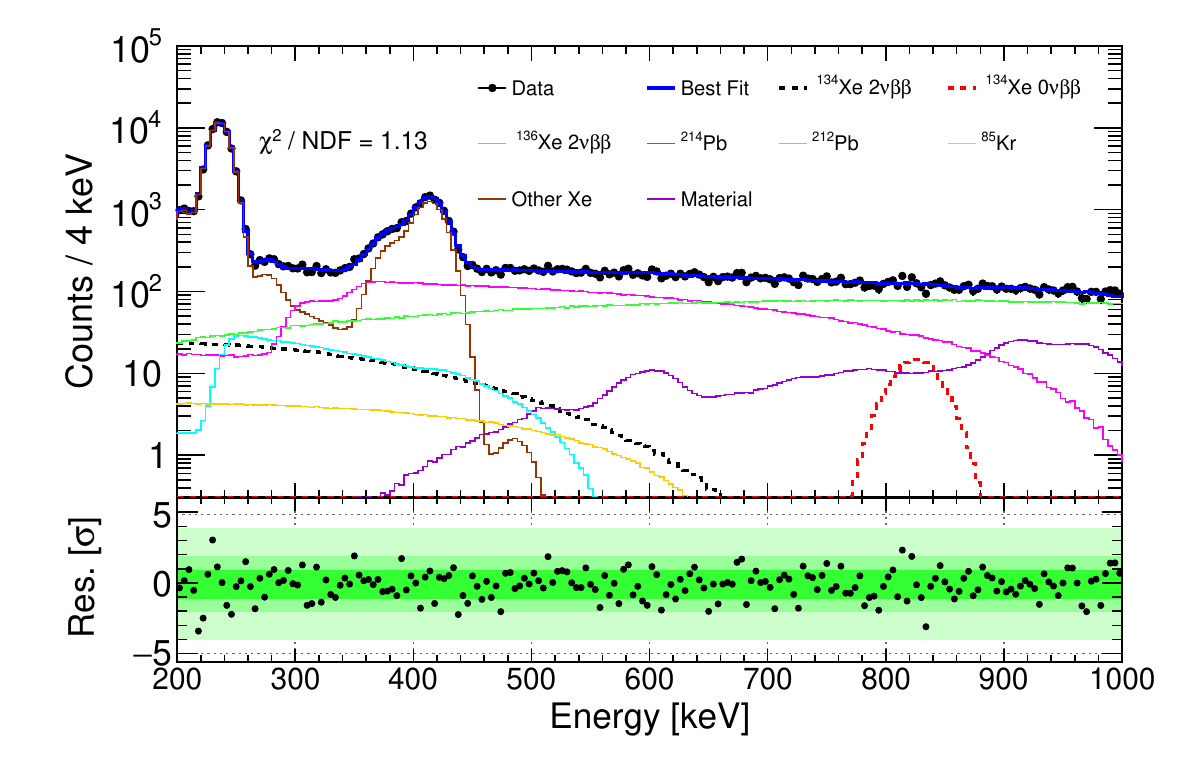}
    \caption{The energy spectrum for the $^{134}$Xe $2\nu\beta\beta$ and $0\nu\beta\beta$ search with fitted background components and the signals at the 90\% CL upper limits. The background contributions of xenon isotopes except $^{136}$Xe are grouped as one curve. 
    The bottom panel shows the residuals.}
    \label{fig:Fit}
\end{figure}

The fit results are shown in Fig.~\ref{fig:Fit}. 
The fitted number of the $^{134}$Xe $2\nu\beta\beta$ and $0\nu\beta\beta$ events are 10 and 105 with statistical uncertainties of 269 and 48.
The fitted contributions from $^{136}$Xe $2\nu\beta\beta$, $^{85}$Kr, and materials are all consistent with the inputs. 
The fitted activity of $^{212}$Pb is $0.30\pm 0.08 ~\mu$Bq/kg, agreeing with the product of the measured $^{212}$Po decay rate and the $^{212}$Pb-to-$^{212}$Po ratio from the calibration. 
The fitted $^{214}$Pb activity is $4.5\pm 0.2 ~\mu$Bq/kg, 63\% of the $^{222}$Rn activity.

The systematic uncertainties are listed in Table~\ref{tb:uncertainty}, which include (1) target mass, (2) efficiency, (3) energy resolution, (4) background, (5) bin size, and (6) energy scale.
The contributions from  1 to 4 are included in the likelihood function as Gaussian constraints [Eq.(\ref{eq:PL1})] and those from 5 and 6 will be evaluated independently.

The uncertainties from target mass include FV, LXe density, and the isotopic abundance of $^{134}$Xe, which are jointly modeled by a common normalization parameter. 
The uncertainties of FV are determined by the $^{\mathrm{83m}}$Kr calibration data.
The observed difference between the geometrically calculated and linearly scaled volume from $^{\mathrm{83m}}$Kr amounts to 1.6\%, representing the uncertainty associated with the FV selection.
The LXe density $2.850\pm 0.004$~g/cm$^3$ is calculated from the gas phase pressure of $2.064 \pm 0.008$ bar during the run period. 
The isotopic abundance of $^{134}$Xe in the PandaX-4T detector is measured with a dedicated setup utilizing an SRS residual gas analyzer~\cite{RGA}.
Considering the relative ionization efficiency of xenon isotopes, the abundance of $^{134}$Xe is calculated to be $10.47\pm 0.02\%$.
The difference with respect to the reported value 10.44\%~\cite{NIST} is treated as a systematic uncertainty. 
Its impact on both $2\nu\beta\beta$ and $0\nu\beta\beta$ is negligible.

Systematic uncertainties from efficiencies include the uncertainty of the SS fraction (6.4\%) and quality cuts (0.03\%), as described previously.
The contribution propagates to an uncertainty of 119 events for $2\nu\beta\beta$ and 10 events for $0\nu\beta\beta$. 

Systematic uncertainties of energy resolution are included by the penalty terms of $a$, $b$, and $c$ in the likelihood function.
The best-fit energy resolution function is consistent within $1\sigma$. 
The change in energy resolution does not introduce noticeable systematic uncertainties to both results.

Systematic uncertainty from background constraints is a dominant term in the final fit results.
Many backgrounds are not well constrained from prior measurements or other energy regions. 
The contributions are 521 and 19 events for $2\nu\beta\beta$ and $0\nu\beta\beta$, respectively.

Additional systematic uncertainties from bin size and energy scale are estimated manually. 
The bin size is varied from 1 to 20~keV. 
The energy scale of PDFs is shifted by up to $\pm10$~keV. 
The maximum changes of best-fit values of $2\nu\beta\beta$ and $0\nu\beta\beta$ are taken as the systematic uncertainties.
The values are listed in Table~\ref{tb:uncertainty}.

The final number of $^{134}$Xe $2\nu\beta\beta$ and $0\nu\beta\beta$ events are $10 \pm 269 (\mathrm{stat.}) \pm 680 (\mathrm{syst.}) $ and $105 \pm 48(\mathrm{stat.}) \pm 38(\mathrm{syst.}) $, respectively. 
The systematic uncertainties are calculated by adding the respective values list in Table~\ref{tb:uncertainty} quadratically.
The $p$ values for null results are found to be 0.99 for the $2\nu\beta\beta$ and 0.09 for $0\nu\beta\beta$, showing that there is no statistically significant evidence for nonzero signals.
The lower half-life limits of 2.8$\times$10$^{22}$~yr and 3.0$\times$10$^{23}$~yr at the 90\% CL are derived, respectively. 

\begin{table}[tbp]
    \caption{Summary of the uncertainties in the number of counts. The fitted mean values of $2\nu\beta\beta$ and $0\nu\beta\beta$ are 10 and 105 counts.} 
    \label{tb:uncertainty}
    \centering
   \begin{tabular}{llll}
   \toprule
   \hline
   \hline
   \multicolumn{1}{c}{}&\multicolumn{1}{c}{Sources} & \multicolumn{1}{c}{~$2\nu\beta\beta$~} &\multicolumn{1}{c}{~$0\nu\beta\beta$}\\ \hline
   \multicolumn{1}{l}{Statistical\quad} &\multicolumn{1}{c}{}& \multicolumn{1}{c}{269} &\multicolumn{1}{c}{48}\\\hline
   \multicolumn{1}{c}{\multirow{6}{*}{Systematic\quad}}&\multicolumn{1}{c}{Target mass} & \multicolumn{1}{c}{-} &\multicolumn{1}{c}{-}\\ 
   \multicolumn{1}{c}{\multirow{6}{*}{}}&\multicolumn{1}{c}{Efficiency} & \multicolumn{1}{c}{119} &\multicolumn{1}{c}{10}\\ 
   \multicolumn{1}{c}{\multirow{6}{*}{}}&\multicolumn{1}{c}{Energy resolution} & \multicolumn{1}{c}{-} &\multicolumn{1}{c}{-}\\ 
    \multicolumn{1}{c}{\multirow{6}{*}{}}&\multicolumn{1}{c}{Background} & \multicolumn{1}{c}{521} &\multicolumn{1}{c}{19}\\
   \multicolumn{1}{c}{\multirow{6}{*}{}}&\multicolumn{1}{c}{Bin size} & \multicolumn{1}{c}{85} &\multicolumn{1}{c}{13}\\ 
   \multicolumn{1}{c}{\multirow{6}{*}{}}&\multicolumn{1}{c}{Energy scale} & \multicolumn{1}{c}{410} &\multicolumn{1}{c}{28}\\ 
    \cline{2-4}
    \multicolumn{1}{c}{}& \multicolumn{1}{c}{Total} & \multicolumn{1}{c}{680} &\multicolumn{1}{c}{38}\\
    \hline
    \hline
   \toprule
   \end{tabular}
\end{table}

In summary, we performed the first experimental search for  $2\nu\beta\beta$ ($0\nu\beta\beta$) of $^{134}$Xe using natural xenon as a target with an isotopic exposure of $17.9 \pm 0.3$~kg yr. 
No significant excess over the expected background is observed, resulting in a 90\% CL lower limit on the half-life of 2.8$\times$10$^{22}$~yr~(3.0$\times$10$^{23}$~yr) for the $2\nu\beta\beta$~($0\nu\beta\beta$) of $^{134}$Xe.
The $2\nu\beta\beta$ limit surpasses the existing limit~\cite{EXO-200:2017vqi} by a factor of 32 and further approaches the theoretical expectation of 3.7--6.09$\times$$10^{24}$ yrs~\cite{Staudt:1990qi,Barros:2014exa}.
The limit for $0\nu\beta\beta$ is 2.7 times stronger than the previous best result in Ref.~\cite{EXO-200:2017vqi}.
Compared to a $^{136}$Xe enriched detector, the searches benefit from the reduced $^{136}$Xe $2\nu\beta\beta$ background.
Thanks to the low energy threshold of PandaX-4T, the fit range is extended down to 200~keV, leading to a further improvement of the $^{134}$Xe $2\nu\beta\beta$ signal over $^{136}$Xe background ratio.  

PandaX-4T continues to take more physics data, and an upgrade plan is developed to optimize the detector performance at the MeV level. 
Larger statistics and higher quality data will further improve understanding of the background model and the search sensitivities.
A discovery of $^{134}$Xe $2\nu\beta\beta$ may be within reach with PandaX-4T and future large xenon TPCs.


 
This project is supported in part by the grants from National Science Foundation of China (No.12105052, No.12090060, No.12005131, No.11905128, No.11925502), by the grants from China Postdoctoral Science Foundation (No.2021M700859, No.2023M744093), and by the grants from Office of Science and Technology, Shanghai Municipal Government (No.18JC1410200, No.22JC1410100). 
We acknowledge support from Double First Class Plan of the Shanghai Jiao Tong University. We also thank the sponsorship from the
Chinese Academy of Sciences Center for Excellence in Particle
Physics (CCEPP), Hongwen Foundation in Hong Kong, and Tencent
Foundation in China. Finally, we thank the CJPL administration and
the Yalong River Hydropower Development Company Ltd. for
indispensable logistical support and other help.

\bibliographystyle{unsrt}
\bibliography{P4Xe134DBD}
\end{document}